\documentclass{article}
\usepackage[utf8]{inputenc}
\usepackage{url}
\usepackage{color}
\usepackage{geometry}
\usepackage{CJK}
\usepackage{indentfirst}
\usepackage{amsmath}
\usepackage{cases}
\usepackage{graphicx}
\usepackage{caption}
\usepackage{subfigure}
\usepackage{multirow}
\usepackage{float}
\usepackage{cleveref}

\usepackage[normalem]{ulem}

\usepackage{authblk}

\title{Improving the performance of reputation evaluating by combining the structure of network and nonlinear recovery}

\author[1,2]{Meng Li}
\author[3,4]{Chengyuan Han}
\author[1,2]{Yuanxiang Jiang}
\author[1,2]{Zengru Di}
\affil[1]{International Academic Center of Complex Systems, Beijing Normal University, Zhuhai, 519087, People’s Republic of China}
\affil[2]{School of Systems Science, Beijing Normal University, Beijing, 100875, People’s Republic of China}
\affil[3]{Forschungszentrum J\"ulich, Institute for Energy and Climate Research - Systems Analysis and Technology Evaluation (IEK-STE), 52428 J\"ulich, Germany}
\affil[4]{Institute for Theoretical Physics, University of Cologne, 50937 K\"oln, Germany}

\date{}

\begin{document}

\maketitle

\begin{abstract} 
Characterizing the reputation of an evaluator is particularly significant for consumer to obtain useful information from online rating systems. Furthermore, to overcome the difficulties with spam attacks on the rating system and to get the reliable on reputation of evaluators is an important topic in the research. We have noticed that most of the existing evaluator reputation evaluation methods only rely on the evaluator’s rating information and abnormal behavior to establish a reputation system, which miss the systematic aspects of the rating systems including the structure of the evaluator-object bipartite network and the effects of nonlinear effects. This study we propose an improved reputation evaluation method by combining the structure of the evaluator-object bipartite network with rating information and introducing penalty and reward factors. This novel method has been empirically analyzed on a large-scale artificial data set and two real data sets. The results show that the proposed method is more accurate and robust in the presence of spam attacks.This fresh idea contributes a new way for building reputation evaluation models in sparse bipartite rating network.

\end{abstract}

\section{Introduction}
The flourishing development of e-commerce is making broad and far-reaching impacts on our daily lives, leading consumers to increasingly rely on the Internet to obtain information about products and services to decide how to consume \cite{muchnik2013social, lu2012recommender, wang2018content}. However, with overwhelming amount of products or services available, potential users are overloaded with information from big data of the quality attributes, performance attributes, and previous reviews \cite{zeng2014information, zhang2012improving}. In order to solve the information overload of users, some e-commerce platforms have implemented online rating systems to help users for information fusion, where evaluators are encouraged to present reasonable ratings for the objects  \cite{liu2017identifying}. These ratings are representations of the inherent quality of objects and reflections of evaluators' credibility. In reality, the current rating systems face many challenges. The unobjective ratings may be given simply because some users unacquainted with the relevant field or their poor judgments \cite{zeng2012removing}. However, unreliable evaluators even deliberately give maximal/minimal ratings for various psychosocial reasons \cite{chung2013betap, yang2013anchoring, zhang2015memory}.
These ubiquitous noises and distorted information purposefully mislead evaluators' choices and decisions and have a wicked effect on the reliability of the online rating systems \cite{toledo2015correcting, zhang2013extracting}. Therefore, establishing a reliable and efficient reputation evaluation system is an extremely urgent task for an online rating system, which has a huge impact not only on against spam attack, but also on the economy and society \cite{allahbakhsh2015robust, dhingra2019spam}.

In various evaluation systems, the reputation management of evaluators contributes to social governance.For instance, as an important platform for providing health services, online health communities are favored by both physicians and patients through establishing effective service channel between them \cite{qiao2021join}.
In the evaluation of research funding applications, peer reviewers must distinguish the best applications from the relatively weaker ones to appropriately allocate funding. Only peer reviewers with a good reputation can correctly guide the highly competitive allocation of limited resources \cite{gallo2016influence, pier2018low}. Moreover, the online reputation system for job seekers helps employers better understand job seekers and decide whether to hire them \cite{novotny2017oblivion}. Similar problems exist in other scenarios, e.g., recommendations, selection, and voting, in which the credibility of the evaluators will affect the final result. One of the most important ways to solve this problem is by building reputation-evaluation systems \cite{fouss2010probabilistic}.
 
Over the past decades, researchers have been interesting increasingly in modeling reputations on web-based rating platforms \cite{wang2014reputation, chang2013building}. The earlier method of measuring the reputation of online evaluators is the iterative refinement (IR) algorithm designed by Laureti \cite{laureti2006information}. The correlation-based ranking (CR) method proposed in \cite{zhou2011robust} by Zhou et al. is the most representative method, and it is robust against spam attacks. Very recently, the IARR2 algorithm was proposed by introducing two penalty factors to improve the CR method  \cite{liao2014ranking}. These aforementioned methods are based on the assumption that each rating given by evaluators is the most objective reflection of the quality of the objects. Another kind of thinking is to consider the behavior features of evaluators in bipartite networks. Gao et al. proposed group-based ranking (GR) and iterative group based ranking (IGR) algorithms. It groups evaluators according to their ratings \cite{gao2015group, gao2017evaluating} and measures the evaluators’ reputation according to the sizes of the corresponding groups \cite{dai2018identifying}.  Other scholars employed the deviation-based ranking (DR) method to model evaluators’ reputation \cite{lee2018deviation}, and Sun et al. combined this method with GR to construct the iterative optimization ranking (IOR) \cite{sun2021evaluating}. In addition, there are some other methods, such as the Bayesian-based method \cite{zhang2013framework, liu2017identifying} and others \cite{liu2015ranking}. One can also read the review literature on the reputation system \cite{lu2016vital} for further research.  

Nevertheless, most existing reputation evaluation algorithms neglect the systematic aspects of the rating systems, especially the structure information of the evaluator-object bipartite network and the effects of nonlinear effect, both of which are core factors in complex systems. 
Considering these factors leads some new ideas to improve the classical CR method. In this paper, we introduce a new reputation evaluation method by combining CR method with the clustering coefficient of evaluators in the evaluator-object bipartite network. Meanwhile, we also believe that if an evaluator has a relatively high reputation, he should receive some rewards to enhance his reputation further, and vice versa. Therefore, we construct a penalty reward function to update the weight of the evaluator's reputation. Extensive experiments on artificial data and two well-known real-world data sets suggest that the proposed method has higher accuracy and recall score of spammer identification. The overall performance exceeds the classical CR method.

The remainder of this paper is organized as follows. The proposed reputation-evaluation method is described in detail in Section \ref{sec:method}. Section \ref{sec:data_and_metrics} introduces the data and evaluation metrics. The experimental study and results are discussed and analyzed in Section \ref{sec:result}. Finally, conclusions are given in Section \ref{sec:conclusion}.

\section{Methods}
\label{sec:method}

We first briefly introduce some basic notations for the online rating systems, which can be naturally represented as a weighted evaluator-object bipartite network. The set of evaluators are denoted $E$, and the set of objects are denoted $O$. The number of evaluators and objects are recorded as $|E|$ and $|O|$ respectively. We use Latin and Greek letters for evaluator-related and object-related indices, respectively. The degree of evaluator $i$ and object $\alpha$ are indicated by $k_i$ and $k_\alpha$. The weight of the link in the bipartite network is the rating given by evaluator $i$ to object $\alpha$, denoted by $r_{i \alpha}$, and $r_{i \alpha} \in [0,1]$. The set $E_\alpha$ describes the evaluators who rate the object $\alpha$, and the set $O_i$ defines the objects rated by the evaluator $i$.

A reputation value $R_i$ should be assigned to each evaluator $i$ by a reputation evaluation method. This value measures the evaluator's ability to reflect the intrinsic quality of the objects or items accurately, known as credibility. 
Similarly, each object $\alpha$ has a true quality that most objectively reflects its character. However, in practice, it is extremely challenging for us to tell the intrinsic quality of an object, and we usually estimate quality $Q_\alpha$ with the weighted average of the ratings that object $\alpha$ has obtained. It is shown as
\begin{equation}
\label{eq:quality}
Q_\alpha=\frac{\sum_{i\in E_\alpha}R_i r_{i\alpha}}{\sum_{i\in E_\alpha}R_i},
\end{equation}
where the initial reputation of each evaluator is set as $R_i=k_i/|O|$. 

Secondly, the CR method defines that the reputation is measured by the correlation between the rating vector from the evaluator and the corresponding quality vectors of the objects. We calculate the evaluator's temporary reputation as
\begin{equation}
\label{eq:temp_reputation}
TR_i=\frac{1}{k_i}\Sigma_{\alpha\in O_i}(\frac{r_{i\alpha}-\overline{r}_i}{\sigma_{r_i}})(\frac{Q_\alpha - \overline{Q}_\alpha}{\sigma_{Q_\alpha}}),
\end{equation}
where $\sigma_{r_i}$ and $\sigma_{Q_\alpha}$ are, respectively, the standard deviations of the rating vector of evaluator $i$ and the corresponding objects' quality vector, and $\overline{r}_i$ and $\overline{Q}_\alpha$ are their mean values. $TR_i$ is reset to 0 if $TR_i$ is less than 0, so that $TR_i$ is limited in the range [0,1]. 

Next, as we mentioned in the introduction, the clustering coefficients in the bipartite graph network are employed to refine the reputation of evaluators. 
Despite the one-mode projection network providing the interaction between each group member, it should be noted that there is substantial information 
may disappear after projection \cite{latapy2008basic}. This paper adopts the concept of clustering coefficient extended by Latapy et al. \cite{latapy2008basic}, who first defines the clustering coefficient for pairs of nodes $cc(e_i, e_j)$. Mathematically, it reads
\begin{equation}
\label{eq:pair_cc}
cc(e_i, e_j) = \frac{|N(e_i)\cap N(e_j)|}{|N(e_i)\cup N(e_j)|}.
\end{equation}
Here, $N(e_i)$ denotes the objects evaluated by evaluator $i$, i.e., the neighbors of node $i$; and $|\cdot|$ denotes the number of the elements in the set. Then, the clustering coefficient for one node is expressed as
\begin{equation}
\label{eq:single_cc}
cc(e_i) = \frac{\sum_{e_j \in N(N(e_i))} cc(e_i, e_j)}{|N(N(e_i))|}
\end{equation}

We now refine the reputation of evaluators according to the clustering coefficient of each evaluator. This modified method is known as CRC 
which can be expressed as follow:
\begin{equation}
\label{eq:mod_temp_rep}
TR'_i = (\frac{cc(e_i)}{\text{max} \{{cc(e_j)}\}})^{\frac{1}{2}}TR_i.
\end{equation}

\begin{figure}[htbp]
    \centering
    \includegraphics[width=0.45\linewidth]{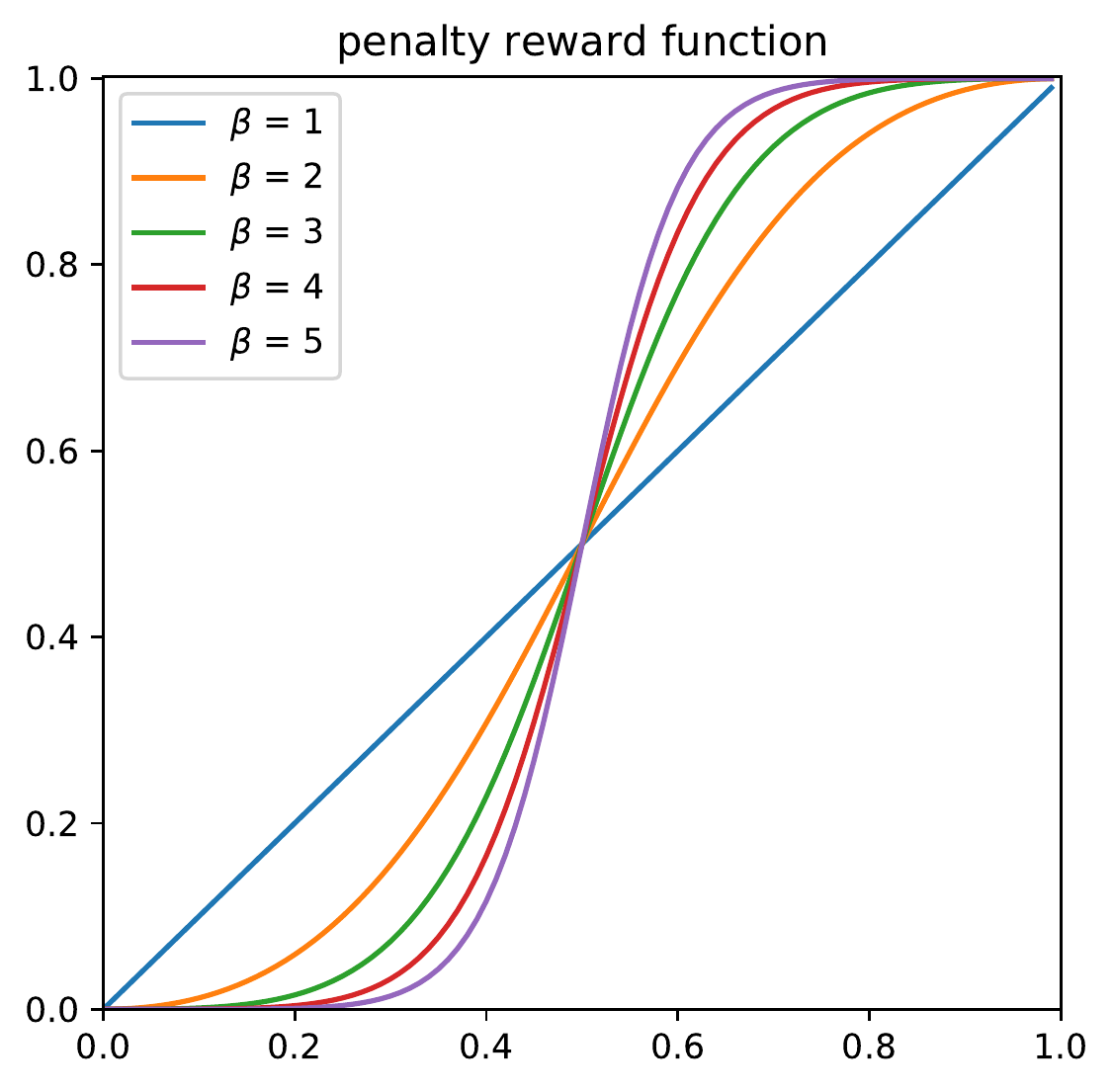}
    \caption{Presentation of the penalty reward function with different parameters $\beta$.}
    \label{fig:beta}
\end{figure}

The second factor we use to update evaluator reputation is the penalty-reward function, which will allocate higher reputations as a reward to evaluators with a high reputation. On the contrary, as a penalty to further reduce the reputations of evaluators with a low reputation. The function is
\begin{equation}
\label{eq:rep}
R_i = \begin{cases}
0 & if \quad TR'_i=0, \\
[1+(\frac{1}{TR'_i}-1)^{\beta}]^{-1} & if \quad 0<TR'_i<1, \\
1 & if \quad TR'_i = 1. 
\end{cases}
\end{equation}
This enhanced method is referred as CRCT and the function image is shown in Figure \ref{fig:beta}. The CRCT method will degrade to the CRC when $\beta=1$.

The evaluator reputation $R_i$ and the quality of the object $Q_\alpha$ are iteratively updated using \cref{eq:quality,eq:temp_reputation,eq:pair_cc,eq:single_cc,eq:mod_temp_rep,eq:rep} until the change of the quality $|Q-Q'|$ is less than the threshold value, and it is calculated in \cref{eq:threshold}. The effects of refining the reputation and estimating the quality are gradually accumulated in each step of the recurring algorithm.  
\begin{equation}
\label{eq:threshold}
|Q-Q'|=\frac{1}{|O|}\Sigma_{\alpha \in O} (Q_\alpha - Q'_\alpha)^2,
\end{equation}
where $Q'$ is the quality from the previous step, and the threshold is set as $10^{-6}$.


Finally, we sort evaluators in ascending order according to their reputation value and the evaluators with $L$ smallest reputation values are identified as spammers. 

\section{Data and metrics}
\label{sec:data_and_metrics}

\subsection{Artificial rating data}
To generate the artificial data set, we generated a biparties network with $6000$ evaluators and $4000$ objects, i.e., $|E| = 6000$, $|O| = 4000$. The network sparsity is set as $\eta=0.02$, which means the total number of the weighted links (ratings) is $0.02×\times|E||O|=4.8×10^5$. We employ the preferential attachment mechanism \cite{barabasi1999emergence} to choose a pair of evaluator and object and add a link between them. At each time step $t$, the probabilities of selecting evaluator $i$ and object $\alpha$ are
\begin{align*}
p_i(t) &= \frac{k_i(t)+1}{\Sigma_{j \in E}(k_j(t)+1)}\\
p_\alpha(t) &= \frac{k_\alpha(t)+1}{\Sigma_{\beta \in O}(k_\beta(t)+1)},
\end{align*}
where $k_i(t)$ and $k_\alpha(t)$ are the degree of evaluator $i$ and object $\alpha$ at time step $t$. 

We suppose that the rating $r_{i \alpha}$ given by evaluator $i$ to object $\alpha$ is composed of the intrinsic quality of object $Q'_\alpha$ and the rating error $\delta_{i\alpha}$. Objects' qualities obey the uniform distribution $U(0, 1)$, and evaluators' rating errors are drawn from the normal distribution $N(0,\delta_i)$. $\delta_i$ indicates the rating error of evaluator $i$, and it is generated from a uniform distribution $U(\delta_{min},\delta_{max})$. In the simulation, we set $\delta_{min}=0.1$ and $\delta_{max}=0.5$. Accordingly, the rating $r_{i\alpha}$ is defined as
\begin{equation}
r_{i\alpha} = Q_\alpha^{'} + \delta_{i \alpha}.
\end{equation}
Both evaluators' ratings and objects’ qualities are limited to the range $[0,1]$.

\subsection{Real rating data}
We consider two commonly studied data sets in real online rating systems---MovieLens and Netflix, which contain ratings for movies provided by GroupLens (\url{www.grouplens.org}) and Netflix Prize (\url{www.netflixprize.com}), respectively---to investigate the effectiveness and accuracy of the proposed methods. These two data sets are given by the integer ratings scaling from 1 to 5, with 1 being the worst and 5 being the best. Herein, we sample a subset from the original data sets in which each evaluator has at least 20 ratings. Table 1 presents some basic statistical properties for these two data sets.

\begin{table}[H]
  \centering
  \setlength{\tabcolsep}{5mm}
  \begin{tabular}{cccccc}
    \hline
    Data set & $|E|$ & $|O|$ & $\left \langle k_e \right \rangle$ & $\left \langle k_o \right \rangle$ & Sparsity \\
    \hline
    MovieLens & 943 & 1682 & 106 & 60 & 0.063 \\
    Netflix & 4960 & 17237 & 295 & 85 & 0.017 \\
    \hline
  \end{tabular}
  \caption{Basic statistical properties of the real data sets used in this paper, where $\left \langle k_u \right \rangle$ and $\left \langle k_o \right \rangle$ are the average degree of evaluators and objects.} 
  \label{Table 1}
\end{table}

It is well known that ranking all evaluators and comparing them with the ground truth is a good way to detect the performance of different evaluation algorithms. However, in real systems, there are no ground-true ranks of evaluators. We manipulate the real data set by randomly selecting some evaluators and assigning them as artificial spammers to test the proportion of these spammers detected by an evaluation method. In the implementation, we randomly select $\rho$ fraction of evaluators and turn them into spammers by replacing their original ratings with distorted ratings: random integers in the set $\{1,2,3,4,5\}$ for random spammers, or integer $1$ or $5$ for malicious spammers. Thus, the number of spammers is $d=\rho|E|$. We also set $\omega=k/|O|$ as the activity of spammers, here $k$ is the degree of each spammer and it is a tunable parameter. If a spammer’s original degree $k_i \geq k$, then $k$ ratings are randomly selected and replaced with distorted ones, and the unselected $k_i-\omega|O|$ ratings are ignored; if $k_i < k$, we first replace all the spammer’s original ratings and randomly select $k-k_i$ of his/her unrated ones and assign them with distorted ratings. 

\subsection{Evaluation metrics}
To evaluate the robustness and the effectiveness of the reputation-evaluation methods, we adopt four widely used metrics: Kendall's tau \cite{kendall1938new}, AUC (the area under the ROC curve) \cite{hanley1982meaning}, recall \cite{herlocker2004evaluating}, and ranking score \cite{zhou2007bipartite}. 

Kendall's tau ($\tau$) measures the rank correlation between the estimated quality of objects $Q$ and their intrinsic quality $Q'$, 
\begin{equation}
\tau=\frac{2}{|O|(|O|-1)}\Sigma_{\alpha < \beta}\mathrm{sgn}[(Q_\alpha-Q_\beta)(Q'_\alpha-Q'_\beta)],
\end{equation}
where $(Q_\alpha-Q_\beta)(Q_\alpha^{'}-Q_\beta^{'})>0$ indicates concordance and $(Q_\alpha-Q_\beta)(Q_\alpha^{'}-Q_\beta^{'})<0$ indicates discordance. Higher $\tau$ values indicate a more accurate measurement of object quality and $\tau \in[-1,1]$.

AUC 
measures the accuracy of the reputation evaluation methods. In artificial data sets, one can select a part of high-quality objects as benchmark objects, and the remaining objects are regarded as non-benchmark 
objects. Here we select $5\%$ of the highest-quality objects as the benchmark objects. Nevertheless, in empirical data sets, as mentioned above, we randomly designate some evaluators as spammers. When the reputation of all evaluators is provided, the AUC value can essentially be interpreted as the probability that the reputation of a randomly chosen normal evaluator is higher than the reputation of a randomly selected spammer. To calculate the AUC values, we control $N$ independent comparisons of the reputations of a pair of normal evaluator and spammer, 
and record $N'$ as the number of times the spammer has a lower reputation, and $N''$ as the number of times they have the same reputation. Then the value of AUC is defined as
\begin{equation}
AUC = \frac{N^{'}+N^{''}}{N}.
\end{equation}
Therefore, the higher the AUC, the more accurate the evaluation method. If the AUC value is 0.5, it indicates that the method is randomly ranked for all evaluators. 

The recall describes the proportion of spammers that can be identified among $L$ evaluators with the lowest reputation. Mathematically, it can be defined as
\begin{equation}
R_c(L) = \frac{d^{'}(L)}{d},
\end{equation}
where $d^{'}(L)$ is the number of detected spammers in the $L$ lowest ranking list, and the range of $R_c$ is $[0, 1]$. Higher $R_c$ indicates a higher accuracy for reputation ranking.

Ranking score (RS) characterizes the effect of evaluation methods by focusing more on the influence of ranking position. Given the ranking of all evaluators, we measure the position of all spammers in the evaluator ranking list. The ranking score is obtained by averaging all spammers, and the specific formula is as follows:
\begin{equation}
RS = \frac{1}{d}\Sigma_{i \in E_s} \frac{l_{i}}{|E|},
\end{equation}
where $l_i$ 
indicates the rank of spammer $i$ in the evaluator ranking list and $E_s$ denotes the set of spammers. Accordingly, RS has the range $[0, 1]$. A good evaluation algorithm is expected to give the spammer a higher rank, which causes a small ranking score. The smaller RS, the higher ranking accuracy, vice versa. 

\section{Results and discussion}
\label{sec:result}
We analyze the performance of the two proposed algorithms for the artificial data set and two commonly studied empirical data sets respectively, and compare them with the classical CR algorithm and IARR2 algorithm. 

\subsection{Results from artificial rating data}
A well-performing evaluation algorithm should defend against any 
distorted information. We first calculate the values of Kendall's tau $\tau$ and AUC on the generated artificial rating data, including spammers, to investigate the robustness of the proposed two methods and the original CR method in protecting against different spammers. We suppose there are two types of distorted ratings: random ratings and malicious ratings. Random ratings mainly come from mischievous evaluators who provide arbitrary and meaningless rating values, and malicious ratings indicate that spammers always give maximum or minimum allowable ratings to push target objects up or down. 

\begin{figure}[ht]
    \centering
    \includegraphics[width=0.98\linewidth]{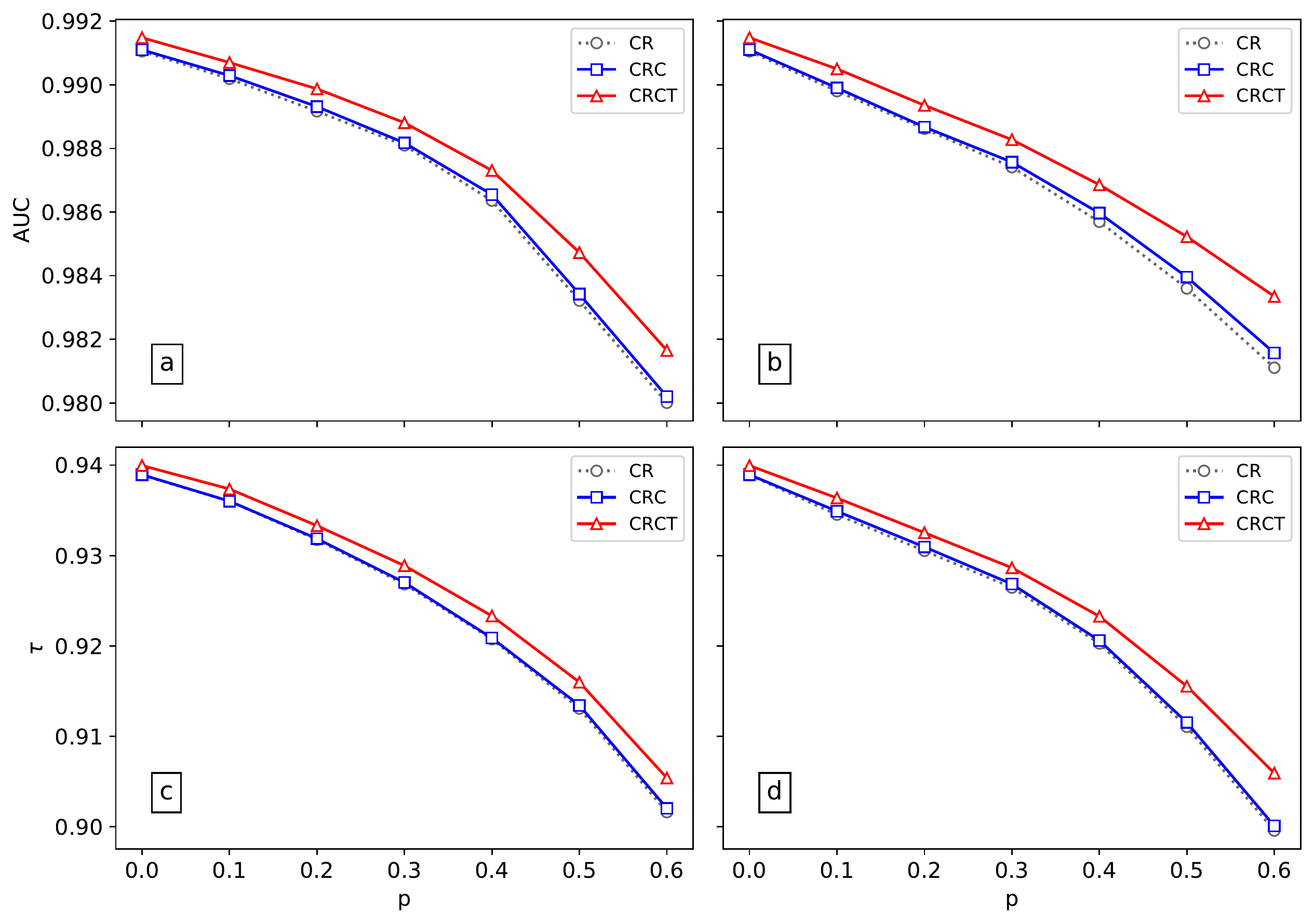}
    \caption{Compare the robustness of the three algorithms. Panels (a) and (c) are the AUC and $\tau$ for different fractions $p$ to random rating spamming, and panels (b) and (d) show the same for malicious rating spamming. The results are averaged over ten independent realizations.}
    \label{Figure 2}
\end{figure}

To create noisy information for the artificial data sets, we randomly switch $p$ fraction of the links with the distorted ratings 
The larger the value of $p$, the less true information in the data set, while $p=1$ means there is no true information. In the following analysis, we set $p\in [0,0.6]$. We report the effectiveness of the two proposed algorithms and the CR method as the ratio of spammers increases. Figure \ref{Figure 2} shows the dependence of AUC and $\tau$ on different values of $p$ for random ratings and malicious ratings. 
For both spammer cases, one can easily observe that the AUC value and $\tau$ of the CRC method are only slightly higher than those of the classical CR algorithm. However, the CRCT method is significantly better than the CR method, especially when the ratio of spammers is high. Thus, we conclude that both of our proposed algorithms, CRC and CRCT, have more advantages than CR method. 

We also investigated the effect of $\beta$ on AUC and $\tau$ in the CRCT method, and the results are shown in the Supplementary Information (SI). 
It is obvious that the parameter $\beta$ improves the effectiveness of the algorithm since CRCT degenerates to the CRC method when $\beta=1$. Moreover, the difference in AUC value between $\beta=2$ and $\beta=3$ is negligible, but $\tau$ is optimal when $\beta=2$, which implies that the overall performance of the CRCT algorithm is better when $\beta=2$. In the following analysis, we adopt $\beta=2$. Please see SI section 1 for the dependence of AUC and $\tau$ on the parameter $\beta$.

\subsection{Results from real rating data}
\begin{figure}[ht]
    \centering
    \includegraphics[width=0.95\linewidth]{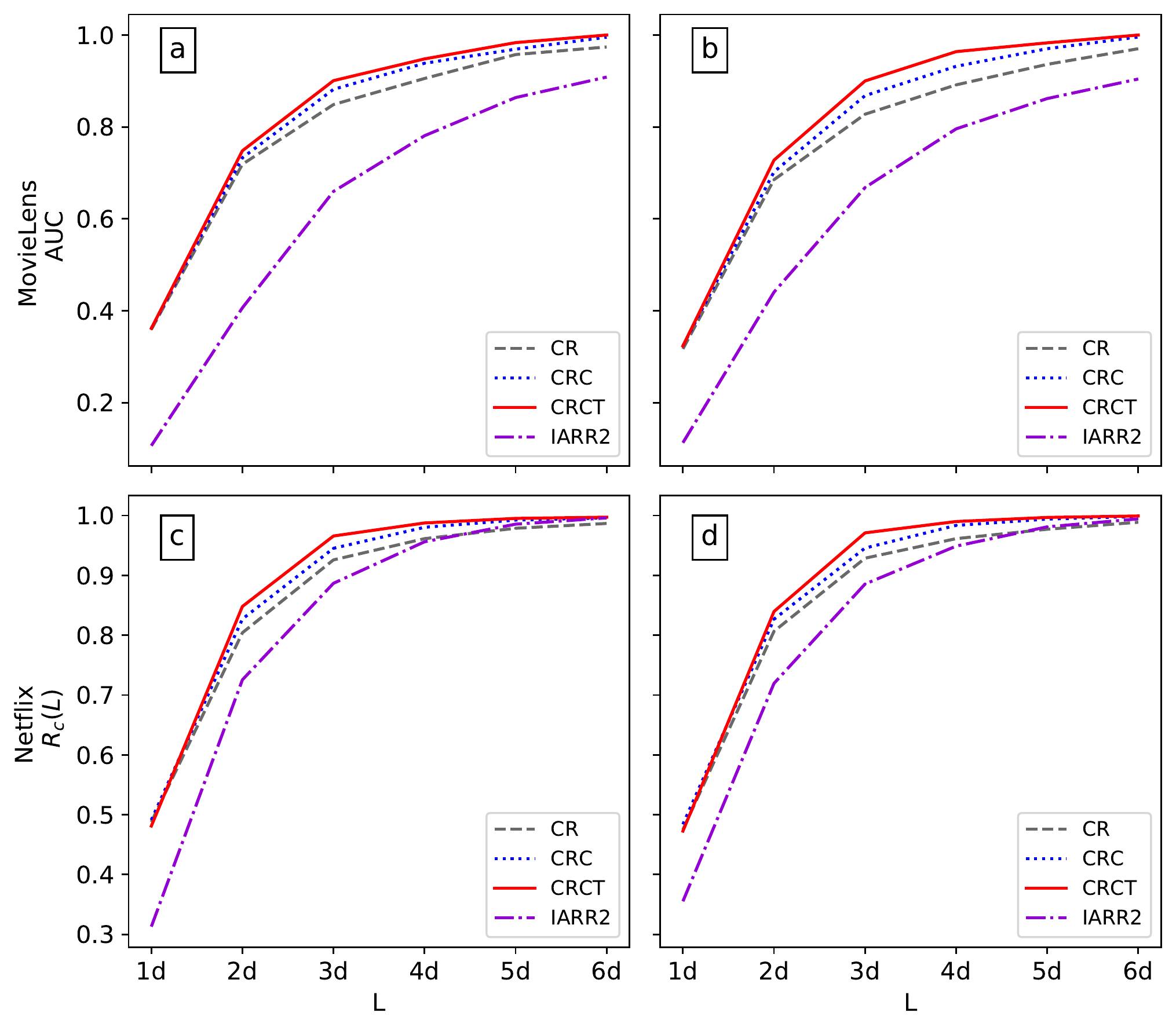}
    \caption{The recall score $R_c$ of different methods varies with length $L$ in MovieLens and Netflix. Panels (a) and (c) stand for random spammers, and panels (b) and (d) stand for malicious spammers. The parameter $\rho$ in both data sets is $0.05$, and the parameter $\omega$ is $0.05$ and $0.01$ for MovieLens and Netflix, respectively. The results are averaged over ten independent realizations.}
    \label{Figure 3}
\end{figure}

We naturally consider the performance of the proposed algorithms on real data sets. The reputation values of all evaluators in each data set are calculated and sorted in ascending order to detect the proportion of the top-$L$ evaluators who are spammers. At the same time, CR and IARR2 methods are compared with the proposed CRC and CRCT methods. We first turn $5\%$ of evaluators in each real data set with two type of spammers to test the effectiveness of the evaluation method, i.e. $\rho = 5\%$. Figure \ref{Figure 3} presents the recall score of different methods calculated according to the length $L$ of the spammer list. We can notice that regardless of the type of spamming, the CRCT method has a significant advantage over the CR method, and the CRC method is essentially an improvement of the CR method. In particular, this enhancement of CRCT is more remarkable for both data sets in the case of malicious spammers, which indicates that it is more challenging to detect random spammers.

The results of AUC and RS values are reported in Table \ref{Table 2}. One can find that for both types of spammers, the AUC values of the CRC and CRCT methods are higher than those of the CR method for every data set, which implies that the two methods have more advantages in accuracy. However, it is worth mentioning that the improvement of the CRC to the CR is very considerable. Moreover, RS verifies the effectiveness of the CRC and the CRCT from another aspect. The smaller the RS, the higher the ranking of spammers. As shown in Table \ref{Table 2}, we easily note that the RS of the CRCT is the smallest for both types of spammers in both data sets. From the above analysis, we can find that the qualitative results of these methods for both types of spammers are very similar, so we will only consider the case of random spammers in the following.
\begin{table}[!h]
    \centering
    \begin{tabular}{ccccccccccc}
    \cline{2-11}
    (a) & \multirow{2}{*}{Data set}  & \multicolumn{4}{c}{AUC} & & \multicolumn{4}{c}{RS} \\
    \cline{3-11}
     & & CR & CRC & CRCT & IARR2 & & CR & CRC & CRCT & IARR2 \\ 
    \cline{2-11}
    & MovieLens & 0.9183 & 0.9236	& \textbf{0.9252} & 0.8664 & & 0.0846 & 0.0795 & \textbf{0.0780} & 0.1460 \\
    & Netflix	& 0.9329 & 0.9383 & \textbf{0.9400} & 0.9239 & & 0.0675 & 0.0624 & \textbf{0.0608} & 0.0801 \\
    \cline{2-11}
    \end{tabular}
    \begin{tabular}{ccccccccccc}
    \cline{2-11} 
    (b) & \multirow{2}{*}{Data set}  & \multicolumn{4}{c}{AUC} & & \multicolumn{4}{c}{RS} \\
    \cline{3-11}
     & & CR & CRC & CRCT & IARR2 & & CR & CRC & CRCT & IARR2 \\ 
    \cline{2-11}
    & MovieLens & 0.9127 & 0.9213 & \textbf{0.9253} & 0.8654 & & 0.0908 & 0.0839 & \textbf{0.0806} & 0.1436 \\
    & Netflix	& 0.9324 & 0.9380 & \textbf{0.9397} & 0.9228 & & 0.0680 & 0.0623 & \textbf{0.0609} & 0.0790 \\
    \cline{2-11}
    \end{tabular}
    \caption{AUC and RS values of different methods on two real data sets for (a) with random spammers and (b) with malicious spammers. The parameters $\omega$ and $\rho$ are the same as Figure \ref{Figure 3}. The results are averaged over ten independent realizations. The most remarkable value in each row is emphasized in bold.}
    \label{Table 2}
\end{table}

\begin{figure}[htbp]
    \centering
    \includegraphics[width=1\linewidth]{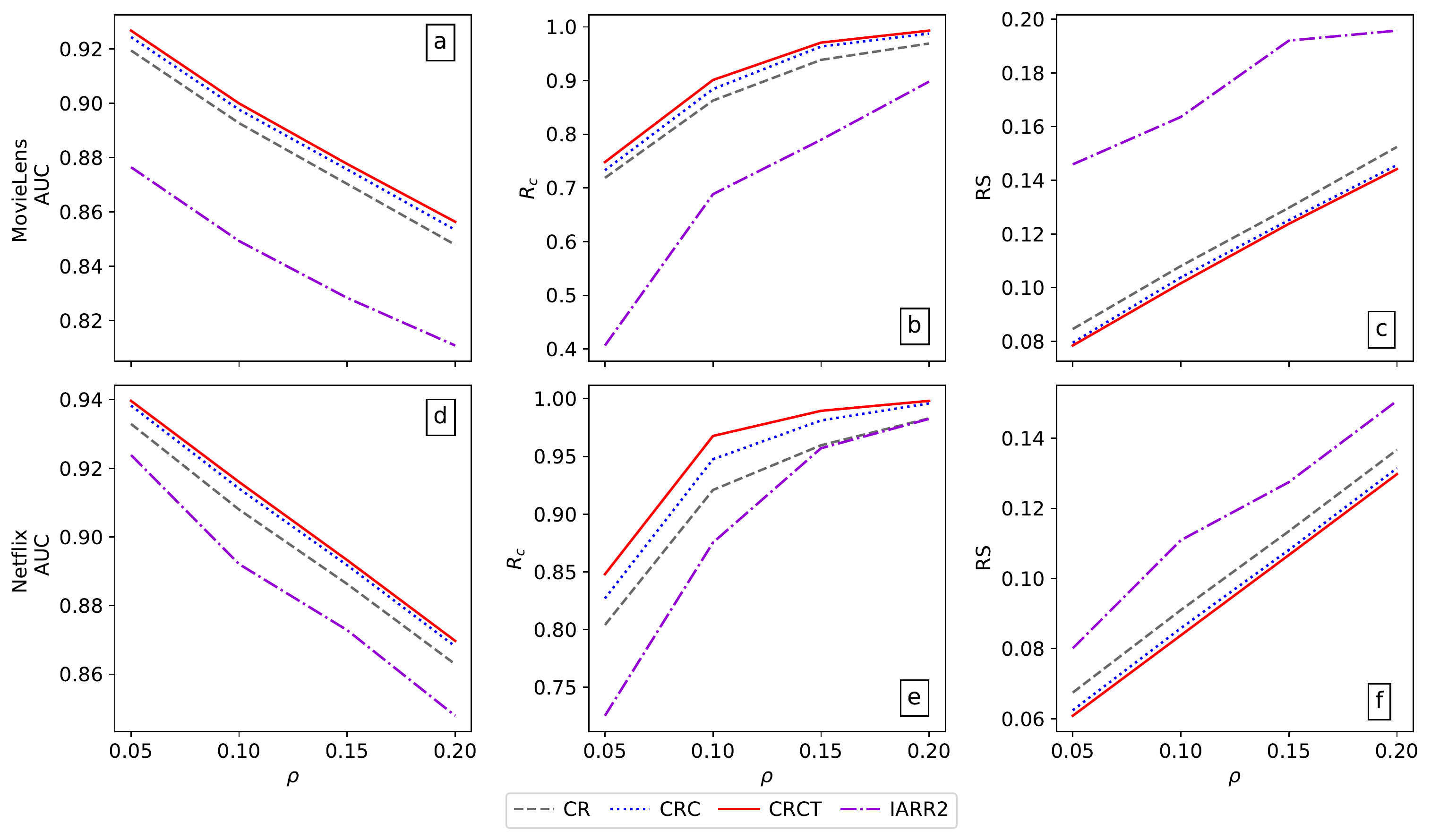}
    \caption{The AUC, $R_c$ and RS values of different methods with different $\rho$ in the random spammers for (a-c) MovieLens and (d-f) Netflix data sets, respectively, and the parameter $\omega$ is $0.05$ and $0.01$ for MovieLens and Netflix. The results are averaged over 10 independent realizations.}
    \label{Figure 4}
\end{figure}

Next, we will analyze whether the performance of the proposed methods is still outstanding while varying $\omega$ and $\rho$, here $\omega$ and $\rho$ are the ratio of objects rated by spammers and the ratio of spammers, respectively. In the following, we set $\rho \in[0.05,0.2]$ to test the robustness changing with the number of spammers in the ground truth and set the length of the detected spam list to twice the number of spammers, namely $L=2d$. The parameter $\omega$ is selected according to the sparsity of the data sets, and $\omega$ of the Netflix data set is smaller than that of the MovieLens data set since the Netflix data set is more sparser. Figure \ref{Figure 4} shows how the AUC, $R_c$, and RS values change under different methods when there are different proportions of spammers in the two data sets. Please see SI section 2 for more details of different $\omega$. It is worth noting that, as a whole, the performance of the CRCT method is better than other methods, 
this is especially when $\rho$ is small. 
Moreover, the $R_c$ of all methods are positively correlated to the number of spammers. In contrast, the RS value of the CRCT method is always lower than other methods, regardless of the number of spammers. Therefore, we conclude that the performance of the proposed CRCT method is stable and accurate.

\begin{figure}[htbp]
    \centering
    \includegraphics[width=1\linewidth]{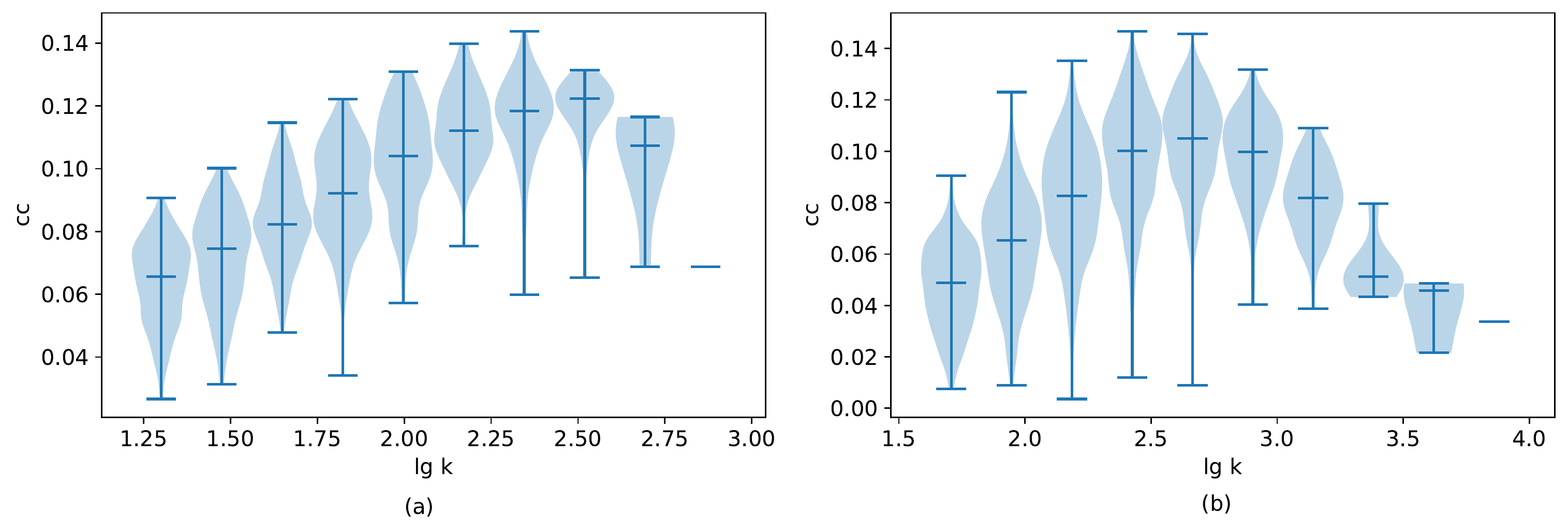}
    \caption{The relationship between the evaluators’ degree and the clustering coefficient in (a) MovieLens and (b) Netflix are presented by the violin plot. The evaluators in each data set are divided into ten bins according to their degrees. The extreme value and median are marked with short bars, and the probability density is represented by shadows.}
    \label{Figure 5}
\end{figure}

One of the motivations of the IARR2 method 
is that evaluators should have a high reputation only when they have a high degree 
However, we can find from Figure \ref{Figure 4} that the IARR2 method performs poorly on the MovieLens data set compared to the Netflix data set, which may be due to the lower average degree of MovieLens data set. On the other hand, this fully demonstrates that the simplest structural information such as degree cannot make a reliable correction to the original CR algorithm. Therefore, it is indispensable to discuss the relationship between the clustering coefficients of evaluators and their degree in the bipartite network, as shown in Figure \ref{Figure 5}. As evaluators' degree are continuous and with different scales, we take the log of the degree for both data sets and divide them into ten bins.
It is not surprising that, like the conclusions of many studies \cite{ravasz2003hierarchical}, there is no relatively positive 
correlation between the evaluators' degree and the clustering coefficient in the two real data sets. To be sure, the introduction of the clustering coefficient in the reputation evaluation process considers the network association from systematic aspects, which effectively improving the classical CR algorithm.

\section{Conclusion}
\label{sec:conclusion}
It is an crucial
problem to build a sound reputation evaluation system for online rating systems. It has great commercial value in e-commerce systems and has guiding significance for a wide range of systematic evaluations. 
In this paper, we propose a robust reputation evaluation algorithm, which considers network association and nonlinear recovery from the systematic aspects of rating system by combining the structural information of the evaluator-object bipartite network and the penalty reward function with the original correlation based ranking method.   
More specifically, in the iterations, we have introduced the clustering coefficient of evaluators in the bipartite network to refine their reputations, then using the penalty-reward function to strengthen the high reputation evaluators further and weaken the impact of the low reputation ones. Extensive experiments on artificial data and two real-world data set show that the proposed CRC and CRCT methods have better performance than the original proposed CR and IARR2 algorithms. These two newly proposed methods outperform the previous ones in evaluating evaluator reputation, and their accuracy and recall scores are remarkably improved and can effectively identify spammers.

The proposed CRCT method has the similar framework as the previous IARR2 algorithm, but the new method focuses more on the core system factors in complex systems and the CRCT method demonstrates its effectiveness and stability over the unsatisfying performance from IARR2.
The results show that introducing the clustering coefficient as the most basic network association feature in the iterative process can capture more profound evaluator behavior characteristics, so as to improve the CR method.
In future work, we can focus on more systematic factors to build a reputation evaluation system, such as the interaction among evaluators. We can also consider the impact of time to build a reputation system because 
normal evaluators 
rarely generate a large number of ratings in a short time, whereas spammers might do so. Additionally, the evolution of evaluator's reputation is also worth exploring in reputation evaluation system.

\section*{Acknowledgement}
This work is supported by the National Natural Science Foundation of China through Grant No.71731002. C.H. gratefully acknowledge support from the German Federal Ministry of Education and Research with grant no.~03EK3055B.

\end{document}


\maketitle

\section{The dependence of parameter $\beta$ in CRCT method}

\begin{figure}[!htbp]
\centering
\includegraphics[width=0.45\linewidth]{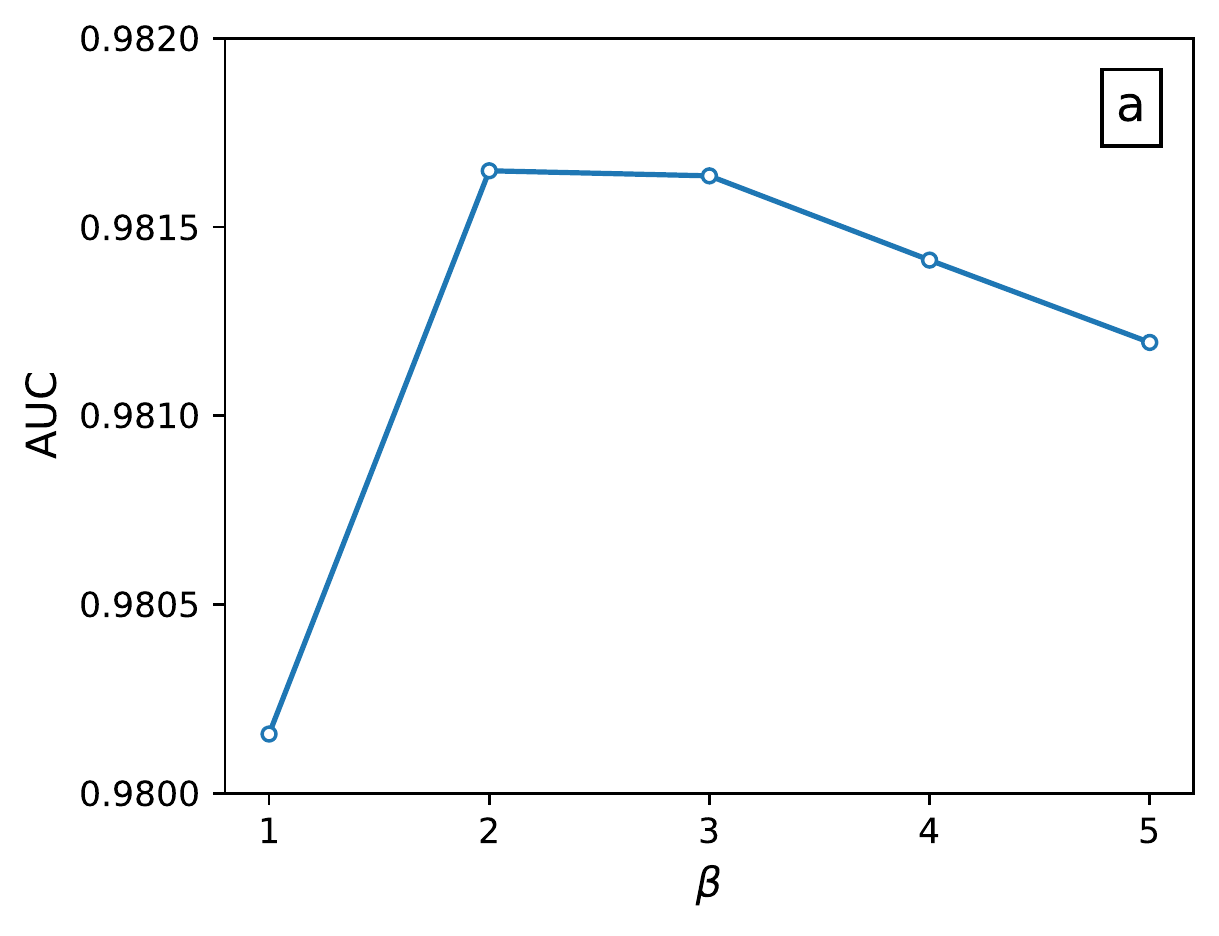} \includegraphics[width=0.45\linewidth]{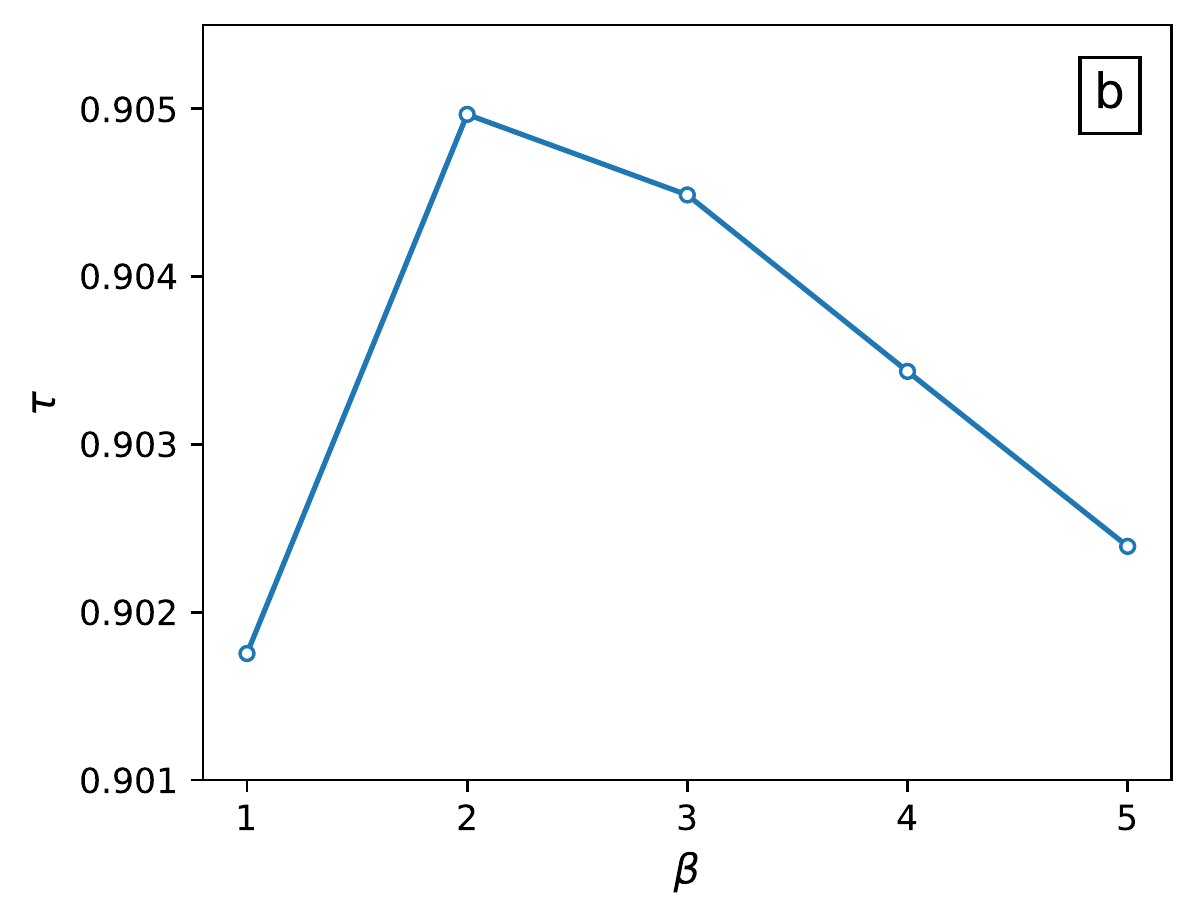} \\
\includegraphics[width=0.45\linewidth]{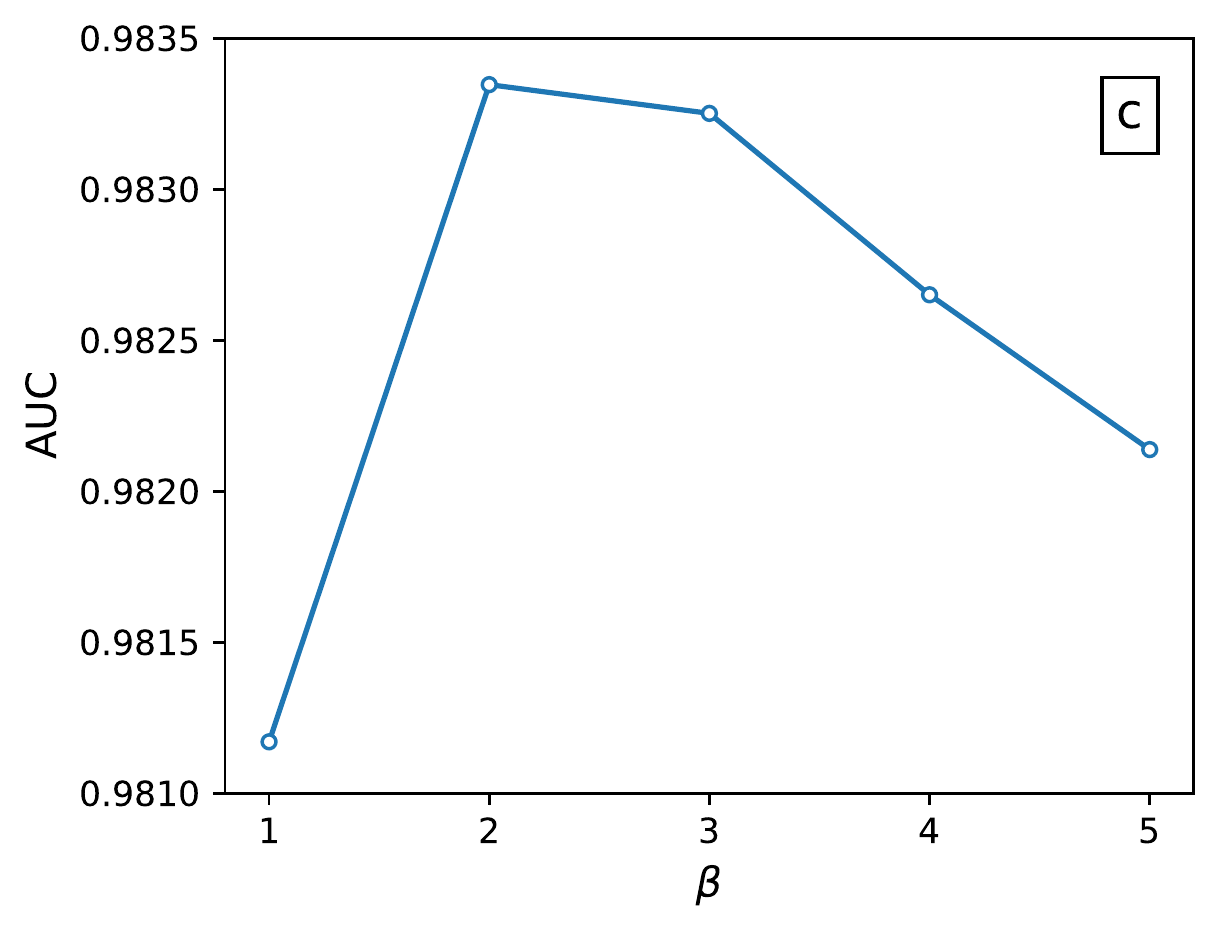}
\includegraphics[width=0.45\linewidth]{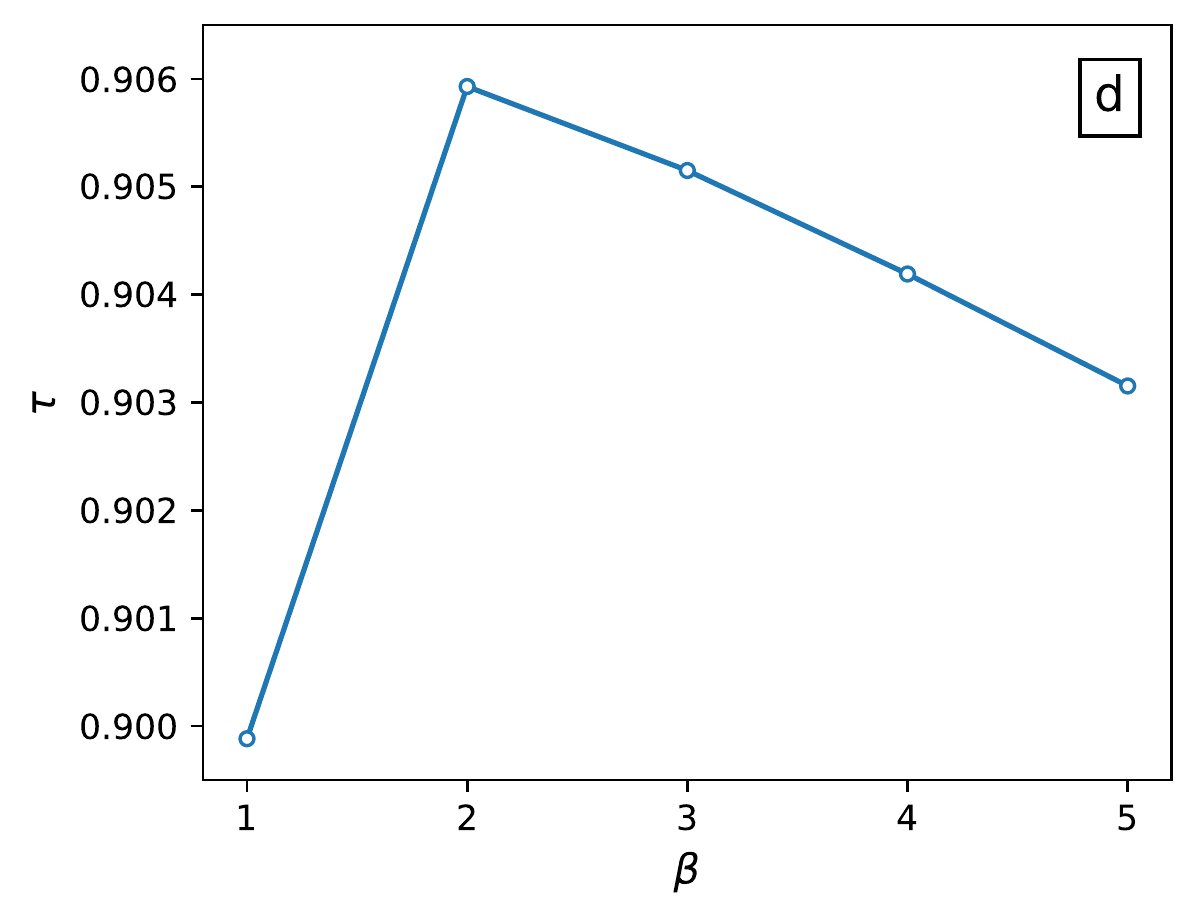}
\caption{The dependence of AUC and $\tau$ on the parameter $\beta$ in CRCT method. Panels (a) and (b) show the AUC and $\tau$ values for different $\beta$ with random-rating spamming, and panels (c) and (d) show the same for push-rating spamming. The results are averaged over ten independent realizations.} 
\label{Figure S1}
\end{figure}

Here, we show the effect of $\beta$ on AUC and $\tau$ in the CRCT method. We set the ratio of spammers to 0.6, and the results are shown in Figure \ref{Figure S1}. As mentioned in the main, the parameter $\beta$ improves the effectiveness of the algorithm since CRCT degenerates to the CRC method when $\beta=1$. Moreover, the difference in AUC value between $\beta=2$ and $\beta=3$ is negligible, but $\tau$ is optimal when $\beta=2$, which implies that the overall performance of the CRCT algorithm is better when $\beta=2$. Therefore, in the following analysis, we adopt $\beta=2$.

\section{Compare the performance of several different methods}
We also compared the performance of the proposed method with the classical CR method and the IARR2 method while varying $\omega$ and $\rho$. We can find that the performance of the CRCT method is better than other methods.
\begin{figure}[h!]
    \centering
    \includegraphics[width=1\linewidth]{SI 2.pdf}
    \caption{The AUC, $R_c$ and RS values of different methods with different $\omega$ and $\rho$ in the random spammers for (a-f) MovieLens and (g-l) Netflix data sets, respectively. Each row represents a different parameter $\omega$: a-c($\omega = 0.75$); d-f($\omega = 0.1$); g-i($\omega = 0.02$); j-l($\omega = 0.03$). The results are averaged over 10 independent realizations.}
    \label{Figure S2}
\end{figure}